\documentclass{aastex}
\usepackage{color}
\usepackage{ulem}
\usepackage{lscape}
\usepackage{graphicx}

\slugcomment{Not to appear in Nonlearned J., 45.}

\shorttitle{
Distance and Proper Motion Measurement 
of S269~IRS~2w
}
\shortauthors{Asaki et al.}

\begin{document}


\title{
DISTANCE AND PROPER MOTION MEASUREMENT \\
OF WATER MASERS IN SHARPLESS~269~IRS~2w
}



\author{Y. Asaki\altaffilmark{1,2}, H. Imai\altaffilmark{3}, A. M. Sobolev\altaffilmark{4}, and S. Yu. Parfenov\altaffilmark{4}}

\affil{$^{1}$
       Institute of Space and Astronautical Science, 
       3-1-1 Yoshinodai, Chuou, Sagamihara, Kanagawa 252-5210, Japan;}
\email{asaki@vsop.isas.jaxa.jp}
\affil{$^{2}$
    Department of Space and Astronautical Science,
    School of Physical Sciences, 
    The Graduate University for Advanced Studies (SOKENDAI),
    3-1-1 Yoshinodai, Chuou, Sagamihara, Kanagawa 252-5210, Japan
}
\affil{$^{3}$
       Department of Physics and Astronomy, 
       Graduate School of Science and Engineering,
       Kagoshima University, 
       1-21-35 Korimoto, Kagoshima 890-0065, Japan;}
\email{hiroimai@sci.kagoshima-u.ac.jp}
\affil{$^{4}$
       Ural Federal University, 
       Lenin Avenue, 51, Ekaterinburg 620000, Russia;
       }
\email{Andrej.Sobolev@urfu.ru, Sergey.Parfenov@urfu.ru}
%
%

\begin{abstract}
We present astrometric analysis of archival data of water masers in 
the star-forming region Sharpless~269 (S269) IRS~2w,  observed with the 
VLBI Exploration of Radio Astrometry. An annual parallax of one 
of the bright maser features in this region was previously reported to be 
$0.189\pm0.008$~milliarcsecond (mas) 
using part of the same archival data as we used. 
However, we found that this maser 
feature is not the best to represent the annual parallax to S269~IRS~2w 
because the morphology is remarkably elongated in the east--west direction. 
For this study we have selected another maser feature showing simpler 
morphology. This makes the new annual parallax estimate more credible.  
Our newly obtained annual parallax is $0.247 \pm 0.034$~mas, 
corresponding to $4.05^{+0.65}_{-0.49}$~kpc. 
This value is well 
consistent with the 3.7--3.8~kpc obtained using the kinematic distance 
estimates and photometric distance modulus. 
We considered two hypotheses for the water maser spatial distribution, 
a bipolar outflow and an expanding ring, in a kinematic model fitting 
analysis with a radially expanding flow. 
At this stage, any conclusions about the systemic proper motion 
could not be drawn from the kinematic analysis. Alternatively, 
we evaluated the mean proper motion to be 
($0.39 \pm 0.92$,~$-1.27 \pm 0.90$)~mas~yr$^{-1}$ 
eastward and northward, respectively, 
from the obtained proper motions of the detected water 
maser features. 
The newly obtained annual parallax and mean proper motion 
give the peculiar motion of S269~IRS~2w to be 
($U_{\mathrm{s}}$,~$V_{\mathrm{s}}$,~$W_{\mathrm{s}}$) of 
($8 \pm 6$, $-21 \pm 17$, $1 \pm 18$)~km~s$^{-1}$. 
\end{abstract}


\keywords{Galaxy: kinematics and dynamics -- Galaxy: structure -- masers -- stars: formation}



\section{
  Introduction
}\label{sec:01}


Sharpless~269 (hereafter, S269) is one of several H~$_{\mathrm{II}}$ 
regions in the outer Galaxy. 
Astrophysical masers in quantum transitions of water molecules 
have been detected in one of the compact infrared sources, 
IRS~2w 
\citep{Lo1973, Genzel1977}.  
%
The water masers were monitored from 2004 to 2006 
with the VLBI Exploration of Radio Astrometry (VERA) of the National 
Astronomical Observatory of Japan (NAOJ) to obtain the annual parallax. 
A sharp-peaked spectrum at $V_{\mathrm{LSR}}$ of 19.7~km~s$^{-1}$ 
was observed at the position of S269~IRS~2w, and the annual parallax 
of the emission was reported to be $0.189 \pm 0.008$~mas 
using model fitting for the eastward sinusoidal motion, 
corresponding to $5.28^{+0.24}_{-0.22}$~kpc 
\citep[][hereafter H2007]{Honma2007}. 

However, this value is considerably higher than distance estimates 
obtained using other methods. 
For example, kinematic distance estimates based on the radial velocity 
of CO molecules provided value of 3.7~kpc 
\citep[see,][] 
{Wouterloot1989}. 
The latest kinematic distance estimated by 
\cite{Xu2009} 
yielded 3.7~kpc using the previously adopted values of galactic rotation 
parameters and 3.0~kpc using the values improved by 
\cite{Reid2009}. 
This estimate was based on the radial velocity of the CS(2--1) molecular 
radio line which traces dense cores associated with H~$_{\mathrm{II}}$ 
regions 
\citep{Bronfman1996}. 
Although kinematic distance estimates are uncertain in general, 
they considerably exceed values of the parallax measurements, 
especially in the outer Galaxy, and S269 was the only  
significant exclusion from this rule 
\citep{Reid2009}. 
Another independent method of the distance estimate based on  
a distance modulus of the luminous star from the S269 stellar 
cluster gives 3.8~kpc 
\citep{Moffat1979}. 
Therefore, H2007's value 
is 40\%--80\% larger than all the above 
distance estimates, and thus there exists an unusually large discrepancy. 

Another aspect which we have to carefully consider for S269 is its 
three-dimensional (3D) motion in the Milky Way. 
H2007 
calculated the 3D motion 
from the absolute proper motion and radial velocity of the 
maser emission. They 
suggested that S269 has a very small peculiar motion with 
respect to a Milky Way flat rotation curve. 
On the other hand,
Miyoshi~et~al. (2012, hereafter M2012) 
showed, 
from the same data as 
H2007, 
that the water maser emissions of S269~IRS~2w are widely 
distributed in space and radial velocity, and have various proper 
motions. 
A systemic proper motion of S269 can be different from the absolute 
proper motion of the single maser emission because it may be a part 
of a complicated internal motion such as an outflow with a typical 
speed of a few tens of km~s$^{-1}$ from a massive protostar. 
It is worthwhile to reanalyze the archival data to intensively inspect 
whether S269 is really in line with the Galactic rotation 
by imaging 
a wide area of the maser emitting region. 

In order to revisit the distance to S269 and its 3D motion, we extensively 
analyzed the VERA archival data, a part of which was already published 
by 
H2007 
and 
M2012, 
including those from the follow-up observations. 
We describe the VERA monitoring program of S269~IRS~2w in 
Section~{\ref{sec:02}}. 
The data reduction procedure is presented in 
Section~{\ref{sec:03}. 
Our astrometric analysis results are described in 
Section~{\ref{sec:04}}. 
We discuss the systemic proper motion and the 3D motion of S269~IRS~2w 
in 
Section~{\ref{sec:05}}, 
and summarize this study in 
Section~{\ref{sec:06}}. 
We often refer results of a specific epoch observation (2005 March 14, or ``epoch C") 
to compare our data reduction process with that previously reported by 
H2007. 
In this paper, 
we adopted a line-of-sight systemic velocity of 
$17.7$~km~s$^{-1}$ 
and its standard deviation of 
$3.6$~km~s$^{-1}$ 
determined from CO molecular line observations  
\citep{Carpenter1990} 
%
%
with the local standard of rest (LSR) defined by 
\cite{Kerr1986}. 
Note that 
we discuss the 3D motion of S269 in 
Section~\ref{sec:05} 
with respect to the solar motion 
reported by 
\cite{Schonrich2010}.  
Hereafter we define a maser ``spot" as an emission in a single 
velocity channel and a maser ``feature" as a group of spots observed 
in at least two consecutive velocity channels at a coincident or 
at very closely located positions
\cite[e.g.,][]{Imai2002}. 

\section{
  Observations
}\label{sec:02}

VLBI phase referencing observations of S269~IRS~2w were conducted 
at 22.2~GHz together with a closely located continuum source, 
ICRF~J061357.6+130645 
\citep[hereafter, abbreviated to J0613+1306,][]{Honma2000},  
$0 \fdg 73$ away, and used as a positional and fringe phase reference. 
Fourteen observations in total were done for two years using 
four VERA antennas.  
The observing epochs are listed in 
Table~\ref{tbl:01}. 
H2007 
and 
M2012 
reported their results using the first six epoch observations. 
%
%
At epochs B and C, there were twelve 16~MHz baseband-converter 
(BBC) channels with a total bandwidth of 192~MHz prepared for 
J0613+1306 while 
a single BBC channel was prepared for S269~IRS~2w. At the other epochs 
there were fourteen BBC channels with a total bandwidth of 224~MHz prepared 
for J0613+1306. 

Observation duration was almost 8.8~hr. Each observation was divided into 
several sessions of 50~minutes each, simultaneously tracking S269~IRS~2w 
and J0613+1306, which were separated by short-term sessions of bright 
calibrators (J022105.5+355613, J041437.8+053442, 3C~120, J052245.1+141529, 
and J053056.4+133155) in order to check the observing systems. 
VLBI cross-correlation 
was carried out 
for the H$_{2}$O $J_{{K}-{K\mathrm{+}}}=6_{16} \rightarrow 5_{23}$ maser line 
with the velocity spacing of 0.2107~km~s$^{-1}$ 
using the Mitaka FX correlator at NAOJ, Mitaka, Japan. 
At epochs~B and C, the central 
4~MHz bandwidth of S269~IRS~2w was cross-correlated while at the other 
epochs the central 8~MHz bandwidth was cross-correlated. 
The cross-correlated fringes were corrected with updated correlator 
models of precise delay tracking supplied by the NAOJ VLBI 
correlation center. 
The positions of the phase tracking centers of S269~IRS~2w and J0613+1306 
were set as listed in 
Table~\ref{tbl:02} 
for all the epochs. In this study we 
express the positions of the water masers in S269~IRS~2w with respect 
to its phase tracking center position. 

\section{
  Data Reduction
}\label{sec:03}
\subsection{
  Preliminary Calibration and Phase Referencing
}\label{sec:03-01}

Data reduction was conducted using the NRAO Astronomical Image Processing 
Software (AIPS) version 31DEC10. 
We first conducted amplitude calibration for all the sources, then 
delay calibrations for S269~IRS~2w and J0613+1306 using the bright 
calibrators. 
Secondly we produced 
a CLEANed image of J0613+1306 for each of the 
epochs using AIPS IMAGR after precise delay, phase, and amplitude 
corrections using AIPS FRING and CALIB. CLEAN components were used to 
calculate J0613+1306's visibility phases of the VERA baselines 
to generate phase-correction data for the phase referencing by 
a direct phase transfer (DPT) method 
\citep{Kusuno2013}. 
The averaged peak flux density for the 14 epochs is 
$0.301 \pm 0.077$~Jy~beam$^{-1}$. We obtained good 
signal-to-noise ratio of the reference source visibilities for 
generating the DPT data.

In order to check the validity of using J0613+1306 as the positional 
reference, we conducted multiple elliptical Gaussian brightness 
distribution fittings of the images using AIPS JMFIT, and found that 
there is a weak component with at most 10\% of the image brightness 
peak, separated 
0.7--1.0~mas from the core component to the north. 
The reference source is an active galactic nucleus (AGN), and the 
brightness peak position may change if relativistic AGN 
jets emerge and/or move. In previous studies with the Very Long 
Baseline Array (VLBA) at 5~GHz, 
multiple jet components to the north can be seen with 
separations of 5--30~mas from the radio core 
\citep{Lazio1998}. 
However, 
another VLBA image of J0613+1306 at 24~GHz shows a compact 
core as well as 
a plausible minor component separated $\sim 1$~mas from the 
brightness peak 
\citep{Charlot2010}, 
which could be identified in our VERA images. 
We cannot find concrete evidence that a new jet component 
was generated in J0613+1306 during this monitoring period. 
J0613+1306's visibility phase calculated from CLEAN 
components of the image was subtracted from the generated 
DPT phase data. 
We conducted phase referencing for S269~IRS~2w using 
the DPT data. 

\subsection{
  Imaging of S269~IRS~2w 
}\label{sec:03-02}

Phase referencing generally reduces short-term phase 
fluctuations of a target source. On the other hand, the synthesis image 
can be more or less distorted by the phase referencing  
because of residual 
errors in the fringe phase on a timescale of several hours. 
To remove such a slow phase change 
from the target source fringe after phase referencing, we used the 
following procedure. 
We first selected a specific velocity channel of S269~IRS~2w in 
which strong maser emissions with a simple structure could be detected. 
Second, we made an initial CLEANed image of the velocity channel. 
Third, we performed fringe-fitting using AIPS FRING for the velocity 
channel with a solution interval of 16~minutes. In this stage, the CLEAN 
components were referred to a source model 
in the fringe-fitting to remove the visibility phase from the FRING solutions. 
We fitted the obtained solutions with third-order polynomials 
and applied the fitted polynomials to all the velocity channels of the 
S269~IRS~2w fringe data to improve the image quality. 
The same polynomials were also applied to the J0613+1306 fringe 
data, and the brightness distribution of J0613+1306 was re-imaged. 
We referred to the peak position of the re-imaged map 
as the positional reference for the astrometry of S269~IRS~2w. 

Note that the absolute positions of the masers are still available after 
this procedure because, if this procedure slightly changes the 
positions of the masers from those of the initial CLEANed image, 
the same positional shift is inevitably induced in the reference 
source by applying the same polynomials. 
This procedure led to not only an improvement of the S269~IRS~2w 
image quality, but also to a distortion of the J0613+1306 image. 
Because we obtained 
a number of maser emissions in the star-forming region while 
there is only one component in the reference source 
this procedure greatly improved efficiencies in determining 
the absolute positions of the maser spots. 
Nevertheless, we recognized large phase offsets 
in AIPS FRING solutions at epochs H to L, and N, which 
distorted the reference source image by applying the fitted polynomials 
and produced clear astrometric outliers. Hereafter, we refer 
to epochs A to G, and M as ``available epochs'' 
for annual parallax analysis, and to epochs H to L, and N as  
``discarded epochs''. 

\subsection{
  Astrometric Accuracy
}\label{sec:03-03}

In the phase referencing a relative position error of the target source with 
respect to the reference source position includes an unexpected peak position 
shift due to inaccuracies in the VLBI correlator model for the pair 
S269~IRS~2w and J0613+1306. Hereafter we refer to the 
relative position error due to the inaccuracies of the VLBI correlator 
model as an 
astrometric error. Let us evaluate the expectation of the astrometric 
error for the 
pair of sources by Monte--Carlo phase 
referencing observation simulations 
using a VLBI observation simulator, 
Astronomical Radio Interferometer Simulator
\citep{Asaki2007, Rioja2012}, 
as described by 
\cite{Asaki2010}. 
In the simulations we assumed the flux densities of S269~IRS~2w and 
J0613+1306 to be equivalently 10 and 0.3~Jy for the 15.6~kHz and 224~MHz 
bandwidths, respectively, which were determined from the observation 
results. 
The sources were assumed to be point sources. The tropospheric zenith 
path length error of 2~cm and other parameters were set as suggested for 
VERA observations 
\citep{Honma2010}. 
The resulting standard deviation of the astrometric error in our observation 
simulations is 
26 and 49~$\mu$as eastward and northward, respectively. These values 
are utilized in the annual parallax analysis and diagnostic morphology error 
estimations described in 
Section~\ref{sec:04-03}. 

\section{
  Results
}\label{sec:04}

\subsection{
  Spatial Distribution of Water Masers of S269~IRS~2w
}\label{sec:04-01}

We produced a 
$16,384 \times 16,384$ 
pixel image cube with a 0.12~mas pixel size  
($1966 \times 1966$ square~mas) 
for the LSR velocity 
($V_{\mathrm{LSR}}$) range of 4--22~km~s$^{-1}$ with a velocity 
channel width of 0.2107~km~s$^{-1}$. 
We conducted elliptical Gaussian brightness distribution 
fitting for each of the detected maser spots using AIPS JMFIT. 
We identified maser spots through the 14 epochs 
whose peak was detected to be brighter by a factor of seven or more 
than the image root-mean-square (rms) noise. Because 
there are a number of spots detected only at a single epoch, 
we selected 90 spots out of them which are detected at least 
at two epochs out of the available epochs, and/or at least 
at three out of all the epochs at the same velocity channel 
in order to make more confident detection. 
We finally identified 28 maser features from the 90 maser spots. 
Figure~\ref{fig:01} 
shows the spatial distribution of the water maser features. 
We labeled the maser feature groups, in which maser features are 
closely located to each other, as groups 1--6 from the most 
northeastern to the most southwestern one, as shown in 
Figure~\ref{fig:01}. 
Table~\ref{tbl:03} 
lists the maser feature group ID, feature ID, radial velocity, 
epoch when the brightness maximum, and the maximum 
brightness value of the 90 maser spots. 

We confirmed that the maser sources are widely distributed within 
a 1~arcsec area as reported by 
M2012. 
Comparing our images with those in 
M2012, the number of detected maser spots is much smaller 
because we excluded image components that were not  
convincingly recognized to be true by visual inspection. 
\cite{Migenes1999} 
found four groups of water maser spots with $V_{\mathrm{LSR}}=16.1$, 
17.3, 19.4, and 20.7~km~s$^{-1}$. We identified the maser groups 
with $V_{\mathrm{LSR}}=19.4$ and 16.1~km~s$^{-1}$ as those involved 
in our maser feature groups~2 and~4, respectively. 
We found that the maser group with $V_{\mathrm{LSR}}=20.7$~km~s$^{-1}$ 
is located 100~mas north of our maser feature 
group~5, but could not identify the maser 
group with $V_{\mathrm{LSR}}=17.3$~km~s$^{-1}$. 

\subsection{
  Individual Images of Water Maser Spots
}\label{sec:04-02}

Figure~\ref{fig:02} 
shows a synthesized image of one of the brightest maser spots with 
$V_{\mathrm{LSR}}=19.7$~km~s$^{-1}$ in maser feature group~2 
(maser spot ID~6 in 
Table~\ref{tbl:03}) 
at epoch~C 
which has been analyzed by 
H2007. 
This maser emission has an extended structure over a few mas 
along the east--west direction. This maser feature may be composed 
of a series of emissions along this direction which cannot be 
distinguished by the synthesized beam. 
Although this maser feature has been very bright through the 
monitoring period, 
it may not be a good idea to use this feature to obtain the annual 
parallax because the positions may have a large uncertainty 
due to such an extended structure. 

Figure~\ref{fig:03} shows two examples of a trial analysis 
to obtain the annual parallax and proper motion for the maser spots. 
At this stage, we used the peak position 
of the brightness distribution for astrometric analysis.
One example shown in 
Figure~\ref{fig:03}(a) 
is a maser spot in maser feature group~3 (maser spot ID~43 in 
Table~\ref{tbl:03}) 
whose synthesis image 
shape is almost the same as the synthesized beam. 
The maser feature including this maser spot shows such 
an image shape, indicating that the maser emission seems to 
have a simple morphology. The obtained annual parallax 
for this single maser spot is $0.239 \pm 0.043$~mas 
by model fitting that simultaneously uses the 
data points in right ascension and declination (combined fitting).
Another example as shown in 
Figure~\ref{fig:03}(b) 
is for the maser spot as shown in 
Figure~\ref{fig:02}. 
As already discussed above, this maser spot can be problematic 
in astrometric analysis because of the structure elongated 
east--west. Because it is difficult to separate positional changes due 
to time variation of an internal brightness distribution along this 
direction from the annual parallax eastward, an expectedly large 
uncertainty in measuring the annual parallax may be induced. 
In 
Section~\ref{sec:04-03}, 
we discuss the positional uncertainty of a maser spot using 
a diagnostic method. 
This example cautions us against using such a maser spot 
showing complicated morphology. 
The obtained annual parallax for this single maser spot is 
$0.134 \pm 0.022$~mas from the combined fitting.

\subsection{
  Annual Parallax
}\label{sec:04-03}

Following several trials of the annual parallax analysis as mentioned above, 
we finalized our analysis to obtain the annual parallax of 
S269~IRS~2w. 
We first selected 
10 from the 90 maser spots (maser spot ID of 42--51), 
(1) detected in at least at five out of the 
available epochs (A--G, and M); 
and 
(2) with the semi-major axis of the fitted elliptical Gaussian 
component whose direction is close to that of the synthesized 
beam within $30^{\circ}$ 
in order to avoid maser spots with complex structures. 
We conducted combined fitting for all the selected maser spots 
simultaneously to a common annual parallax, the positions 
at epoch~A without the annual parallax modulation 
(initial position), and proper motions of the individual maser 
spots. Details of the combined fitting are described by 
\cite{Kusuno2013}. 

We iteratively checked the residuals of the positions 
after subtracting the common annual parallax and individual proper 
motions from the spatial motions of the maser spots. 
The weighted rms of the residuals are 36 and 41~$\mu$as 
eastward and northward, respectively. 
Let us estimate the positional uncertainty due to the maser spot 
morphology (morphology error) using the following diagnostic analysis. 
As mentioned in 
Section~\ref{sec:03-03}, 
the astrometric error is expected to be 26 and 49~$\mu$as eastward 
and northward, respectively. Therefore, the position error due mainly to 
the morphology error is $\sqrt{36^{2}-26^{2}}=25$~$\mu$as eastward. 
On the other hand, the morphology error in declination could not be 
calculated because $41^{2}-49^{2} < 0$. We adopted the morphology 
error of 25~$\mu$as for the maser spots of S269~IRS~2w. 
The morphology and astrometric errors were added in root-square-sum 
(RSS) to each of the components of the positional errors of the water 
maser spots eastward and northward. 
Note that those 
additional errors make the $\chi^{2}$ per degree of freedom to be close 
to unity. We obtained the annual parallax to be $0.247 \pm 0.011$~mas. 
The spatial motions of all the selected ten maser spots are shown in 
Figure~\ref{fig:04} 
with the common annual parallax of 0.247~mas.

We note that all the spots are selected from the specific 
maser feature 3-k. In the last stage, 
the estimated error obtained in our annual parallax analysis for the 
selected maser spots should be multiplied by a factor of $\sqrt{N}$, 
where $N$ is the number of selected maser spots ($N=10$). 
The resultant annual parallax error is 0.034~mas, 
corresponding to 
$4.05^{+0.65}_{-0.49}$~kpc. 
Because this annual parallax is consistent with 
the kinematic distance 
estimates of 3.7~kpc 
\citep[e.g.,][]{
Wouterloot1989, Xu2009} 
and 
the photometric distance modulus of 3.8~kpc 
\citep{Moffat1979}, 
a large discrepancy between the annual parallax and the other 
distance estimates for S269 is eliminated. 

For information, we present the annual parallax of the maser feature 
2-a, which has been analyzed by  
H2007. 
Note that this maser feature was not included in our annual parallax 
analysis because of its problematic morphology. 
We selected the three consecutive velocity channels of the brightest 
maser spot centered at $V_{\mathrm{LSR}}=19.7$~km~s$^{-1}$ as 
H2007 
did. We applied the combined fitting to the three 
spots eastward and northward simultaneously while  
H2007 
made their model fitting only of the right ascension for each of the maser spots, 
then averaged the three results. We obtained the annual parallax to be 
$0.193 \pm 0.031$~mas, 
corresponding to $5.19^{+0.98}_{-0.82}$~kpc, consistent with that of 
H2007. 
Note that the above uncertainty is a product of the combined fitting 
error and a factor of $\sqrt{N}$, where $N=3$ in this case. 
With the diagnostic analysis mentioned above, the morphology 
error of the three maser spots was estimated to be 
43 and 29~$\mu$as eastward and northward, respectively, for 
the annual parallax of 0.247~mas. 

\subsection{
  Absolute Proper Motions
}\label{sec:04-04}

We made use of all 14 epochs to obtain absolute proper motions of the maser 
spots. In this analysis, we have to align the maser spots detected at the discarded 
epochs with the maser spot map of the available epochs. 
In the first step, the annual parallax modulation of 0.247~mas was subtracted 
from the maser positions at the available epochs. 
We then calculated the absolute positions of maser spot ID~6 
(the brightest spot in maser feature 2-a) 
at the discarded epochs from the preliminarily 
obtained initial position 
and the proper motion at the available epochs in 
Table~\ref{tbl:03} 
In the third step, 
the maser spot positions at the discarded 
epochs were determined in the map of the available epochs 
with respect to the calculated position of maser spot ID~6. 
We set the morphology error of all the maser spots 
at the discarded epochs to 
$\sqrt{25^{2}+43^{2}}=50$~$\mu$as and 
$\sqrt{25^{2}+29^{2}}=38$~$\mu$as eastward and northward, respectively, 
and added these to each of the positional errors in RSS. 
In the last step, we obtained the initial positions and absolute proper 
motions eastward and northward, respectively, with least square 
fitting for each of the 90 maser spots. The obtained parameters 
of the selected maser spots are listed in 
Table~\ref{tbl:03}. 
Figure~\ref{fig:01} 
depicts the proper motions and radial LSR velocities of 28 maser features 
by averaging the proper motions of the maser spots identified to be the 
same maser feature with a weighting of the maximum brightness 
as listed in 
Table~\ref{tbl:03}. 

\section{
  Discussions
}\label{sec:05}

The mean motion of the obtained proper motions, 
$\overline{\mbox{\boldmath $M$}}$, shows biased kinematics 
of ($0.39 \pm 0.92$,~$-1.27 \pm 0.90$)~mas~yr$^{-1}$, 
eastward and northward, respectively. To investigate a systemic 
proper motion of S269~IRS~2w, 
$\mbox{\boldmath $M$}_{\mathrm{0}}$, we conducted kinematic 
model fitting analysis for the absolute proper motions of the water 
maser features  
by assuming a radially expanding flow from a single 
originating point, 
$\mbox{\boldmath $x$}_{0}$, 
indicating the position of the massive young stellar object (YSO; 
Imai~et~al. 2011). 
We adopted the Levenburg--Marquart algorithm to search for the solutions. 
An additional proper motion error of 10~mas~yr$^{-1}$ was 
added to each of proper motion components eastward and northward 
in an RSS in order to make the $\chi^{2}$ per degree of freedom 
almost unity. 

However, we found that the resultanting solutions are highly dependent 
on initial values of the parameters in the search. Because the location 
of the dynamical center is uncertain, we therefore consider constraints on 
the position of the originating point. 
Because the water masers in S269~IRS~2w form aligned structure, there 
are two distinguishable major cases: with the dynamical center aligned with 
the masers, and with the dynamical center on the side of the maser structure. 
These two cases have both statistical and physical explanations, 
which we will describe below. 

In the statistical sense it is natural to consider locations of the dynamical 
center that correspond to convergence solutions. The primary candidate 
in this case is the best convergence solution, which is realized at 
$\mbox{\boldmath $x$}_{0}=$~($400$,~$180$)~mas. 
We dare to mention that this position is close to the mean position of 
the water maser spatial distribution and the near-infrared (NIR) 
$K$ band Two Micron All Sky Survey 
(2MASS) point source 
\citep{Skrutskie2006} 
2MASS~J06143706+1349364 
by chance, as shown in 
Figure~\ref{fig:01}. 
This 2MASS point source has an accuracy of 80~mas in the FK5 frame. 
Considering the 
difference between the FK5 frame and the ICRF, the relative positional 
accuracy of the point source is estimated to be $\sim 150$~mas 
at worst in the water maser map. 
The NIR source appears only 
in the $K$ image and is not pronounced in the $J$ and $H$ colors. This 
means that the source is a deeply embedded YSO. 
This case suggests that the water masers 
are associated with a collimated bipolar outflow. 
A piece of evidence that the water masers trace the collimated 
outflow is found in the global structure around S269~IRS~2w. 
\cite{Eiroa1994} 
suggested that a Herbig--Haro object, HH~191, which is located $2 \farcs 8$ to 
northeast of S269~IRS~2w is plausibly produced in a bow shock  
in a flow from S269~IRS~2w. The alignment of the water maser features 
is quite similar to the direction between S269~IRS~2w and HH~191. 
In the case 
with the first candidate of $\mbox{\boldmath $x$}_{0}$, 
we obtained 
$\mbox{\boldmath $M$}_{\mathrm{0}}=$~($0.78 \pm 0.13$,~$-1.06 \pm 0.15$) mas~yr$^{-1}$. 

Another candidate was selected from one of the most convergent 
solutions, 
which is situated beside the maser structure at 
$\mbox{\boldmath $x$}_{0}=$~($-286$,~$515$)~mas. 
This case corresponds to a situation different from the association with 
the collimated outflow. One such possibility is that S269~IRS~2w's 
water masers may trace the cavity walls of a one-sided outflow with a 
rather wide opening angle, as observed in G35--0.74  
\citep{DeBuizer2006}. 
Alternatively, the spatial distribution of the water masers may trace 
circumstellar structures of other nature, e.g., 
they can be associated with an expanding bubble 
\citep{Torrelles2001} 
or a shock wave propagating into a rotating disk 
\citep{Gallimore2003}.   
In the case of the second candidate 
we obtained  
$\mbox{\boldmath $M$}_{\mathrm{0}}=$~($-0.22 \pm 0.14$,~$-0.54 \pm 0.14$) mas~yr$^{-1}$. 

At this stage, we do not have any concrete evidence of the location 
of the young star. In the following discussions, we obtain the 3D motion 
of S269~IRS~2w in the Milky Way using its mean proper motion.  
This proper motion shows a discrepancy with that from 
H2007 
because they showed the analysis results only for maser feature 2-a. 
If we adopt the Galactocentric distance to the Sun of 
$8.3 \pm 0.2$~kpc 
and a flat Galactic rotation curve with rotation velocity of 
$239 \pm 7$~km~s$^{-1}$ 
\citep{Brunthaler2010}, 
the peculiar motion of S269~IRS~2w is 
($U_{\mathrm{s}}$,~$V_{\mathrm{s}}$,~$W_{\mathrm{s}}$) $=$  
($8 \pm 6$, $-21 \pm 17$, $1 \pm 18$)~km~s$^{-1}$, 
where 
$U_{\mathrm{s}}$ is toward the Galactic center, 
$V_{\mathrm{s}}$ is toward the Galactic rotation direction, and 
$W_{\mathrm{s}}$ is toward the Galactic north pole at the position 
of the object. 
For the solar motion, 
we adopted the latest value obtained by 
\cite{Schonrich2010}. 
The peculiar motion of S269~IRS~2w is prominent in the $V_{\mathrm{s}}$ 
component though we note that there 
is a rather large uncertainty because of the error in the systemic 
proper motion, especially in the $V_{\mathrm{s}}$ and $W_{\mathrm{s}}$ 
components in the case of the direction to S269 at a galactic longitude 
of $196 \fdg 5$. 

\section{
  Conclusions
}\label{sec:06}

We presented the data reduction results of the VERA archival data 
of S269~IRS~2w, part of which was already analyzed by 
H2007 
and 
M2012. The water masers ranging in LSR velocity from 4 to 
21~km~s$^{-1}$ spatially distributed along the northeast--southwest direction. 
We obtained the annual parallax of $0.247 \pm 0.034$~mas for 
a maser feature with simple morphology residing in this 
star-forming region, 
corresponding to $4.05^{+0.65}_{-0.49}$~kpc. 

In order to consider the systemic proper motion of S269~IRS~2w we 
presented two major hypotheses regarding the dynamical center position: 
within the aligned maser structure and beside it. 
We found that at present these two cases cannot be distinguished 
by reasoning based on both statistical and physical plausibility. 
Because we have not had clear evidence on the certain location 
of the source of maser excitation, we found it impossible to draw 
any conclusion about the systemic proper motion. 
Instead, 
using the mean motion of the water maser proper motion, 
($0.39 \pm 0.92$,~$-1.27 \pm 0.90$) mas~yr$^{-1}$, the annual 
parallax of $0.247 \pm 0.034$~mas, 
and radial velocity of the S269 molecular cloud, 
the peculiar motion of S269~IRS~2w is estimated to be 
($U_{\mathrm{s}}$,~$V_{\mathrm{s}}$,~$W_{\mathrm{s}}$) of 
($8 \pm 6$, $-21 \pm 17$, $1 \pm 18$)~km~s$^{-1}$. 



\acknowledgments

The authors thank an anonymous referee for careful reviews and 
providing valuable suggestions to improve this paper. 
The VERA/Mizusawa VLBI observatory is a branch of the National 
Astronomical Observatory of Japan, National Institutes of Nature Sciences. 
The authors express their deep gratitude to the VERA of NAOJ. 
The authors acknowledge arrangements for providing the updated 
VLBI correlator models by K.~M.~Shibata   
and fruitful discussions on the VERA data analysis with T.~Oyama. 
%
%
This publication also makes use of data products from the 2MASS, 
which is a joint project of the University of Massachusetts and the 
Infrared Processing and Analysis Center/California Institute of Technology, 
funded by the National Aeronautics and Space Administration and the 
National Science Foundation.
%
%
This work was supported by the Japan Society for Promotion of Science 
(JSPS) Grant-in-Aid for 
Challenging Exploratory Research, grant no. 25610043. 
Y.~A was financially supported by the Center for the Promotion 
of Integrated Science of the Graduate University of Advanced 
Studies for this publication. 
\clearpage
\onecolumn
\begin{figure}[htbp]
\centering
\begin{tabular}{c}
\includegraphics[width=12cm]{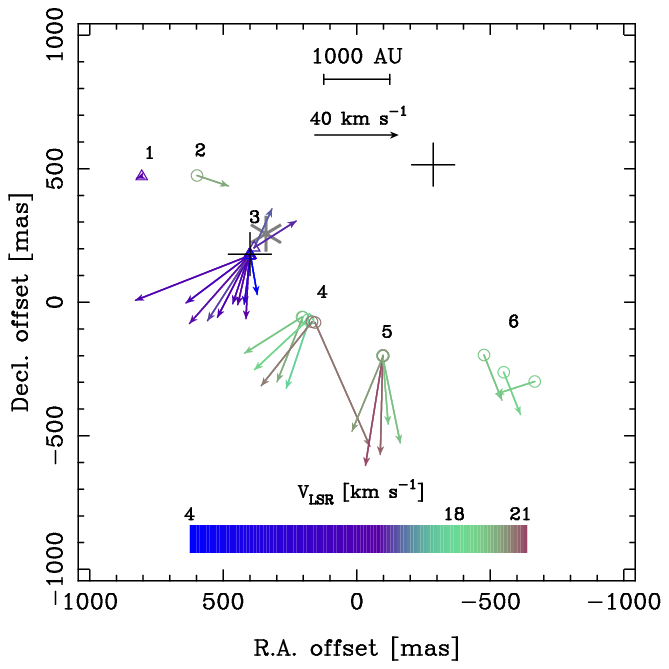}
\end{tabular}
\caption
{\label{fig:01}
  Spatial distribution and absolute proper motions of the water 
  maser features in S269~IRS~2w. 
  Open circles and triangles represent 
  red- and blueshifted features with respect to the systemic velocity 
  of S269. 
  The numbers in the plot represent the maser feature groups identified 
  by the authors. 
  The crosses show the assumed dynamical center discussed in 
  Section~\ref{sec:05}. 
  The gray asterisk represents the position of the 2MASS point source 
  2MASS~J06143706+1349364. 
}
\end{figure} 
\clearpage
%
%
\begin{figure}
\begin{center}
\begin{tabular}{c}
\includegraphics[width=8.0cm]{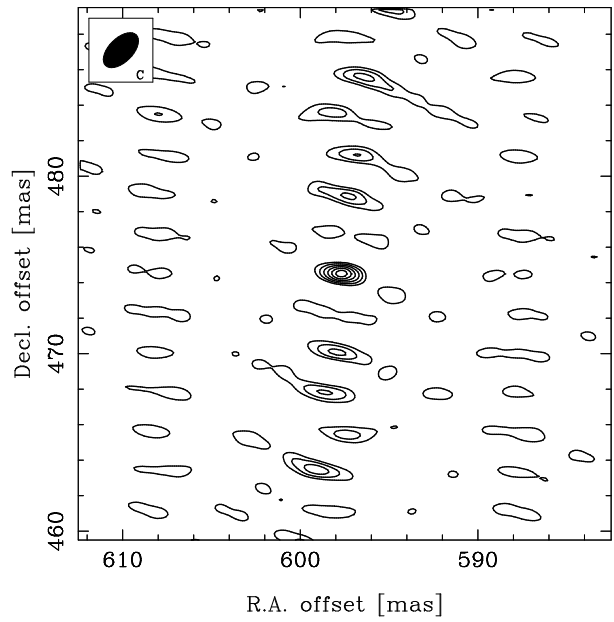}
\end{tabular}
\end{center}
\caption 
{\label{fig:02}
Brightness distribution of S269~IRS~2w around 
the brightest peak with 
$V_{\mathrm{LSR}}=19.7$~km~s$^{-1}$ at epoch C. The contour levels 
are linearly increased by 15\% of the peak strength (98~Jy~beam$^{-1}$). 
The synthesized beam is shown 
in the upper $1 \times 1$ square mas box. 
}
\end{figure} 
%
%
\begin{figure}
\begin{center}
\begin{tabular}{c}
\includegraphics[width=15.0cm]{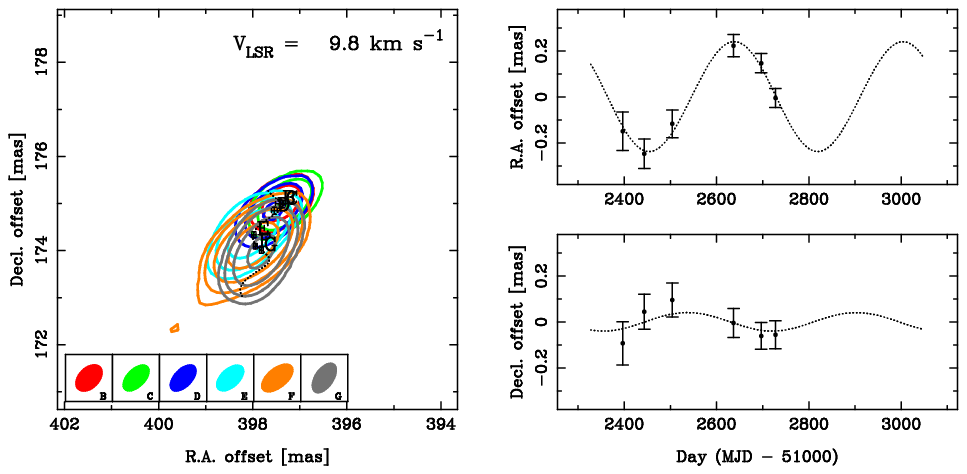} \\
\includegraphics[width=15.0cm]{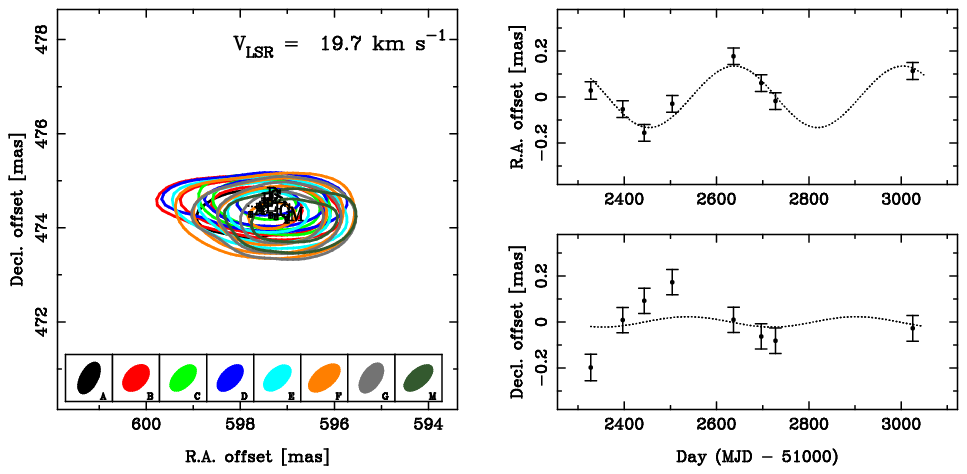}
\end{tabular}
\end{center}
\caption 
{\label{fig:03}
Astrometric analysis results of 
water maser spots in S269~IRS~2w. The top three panels show the 
maser spot with $V_{\mathrm{LSR}}=9.8$~km~s$^{-1}$ in 
maser feature group~3. 
The left panel shows the image of the spots colorized for the observing 
epochs. The outermost contour is the 5$\sigma$ noise level. The 
synthesized beams are shown in the bottom of the plot by $1 \times 1$ square mas. 
The two upper right plots show the maser spot motion eastward and northward 
after removing the obtained proper motion. 
The bottom three panels show the 
maser spot with $V_{\mathrm{LSR}}=19.7$~km~s$^{-1}$ in maser feature group~2. 
The outermost contour shows the 40$\sigma$ noise level. 
The obtained annual parallaxes for the 9.8 and 19.7~km~s$^{-1}$ 
components are 
$0.239 \pm 0.043$
and 
$0.134 \pm 0.022$~mas, respectively. 
}
\end{figure} 
%
%
\begin{figure}
\begin{center}
\begin{tabular}{c}
\includegraphics[width=75mm]{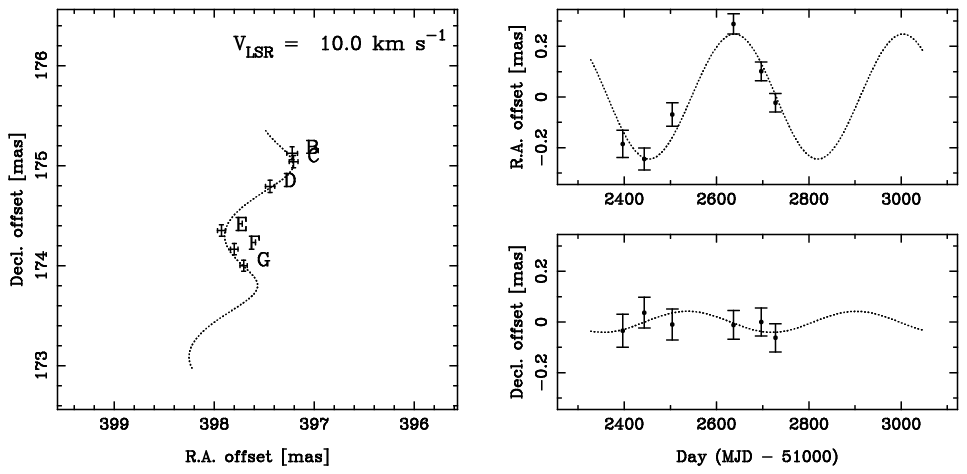}
\includegraphics[width=75mm]{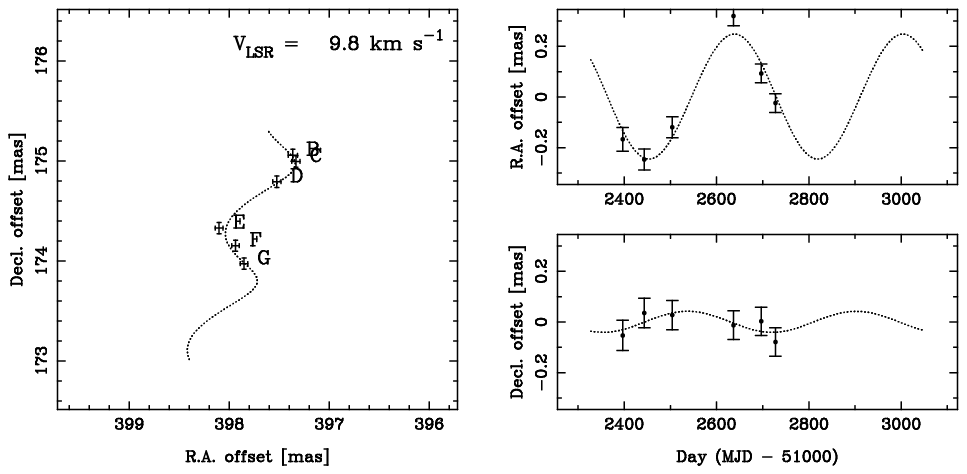} \\
\includegraphics[width=75mm]{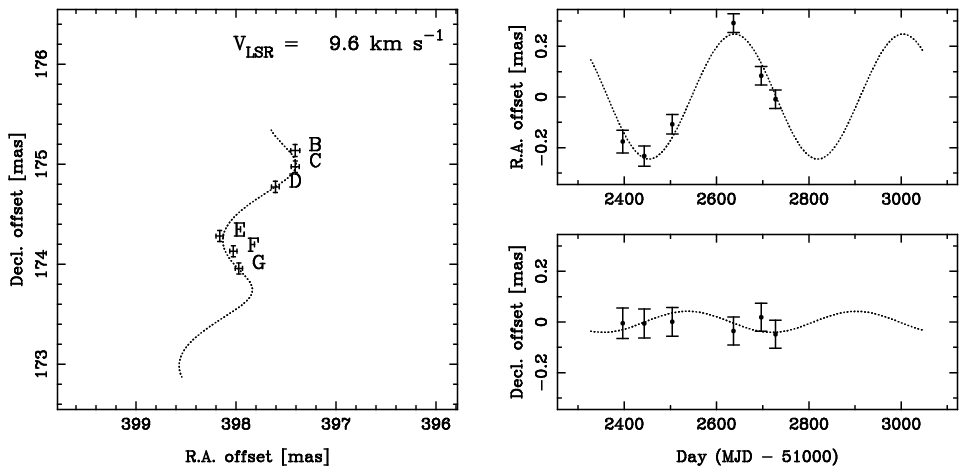}
\includegraphics[width=75mm]{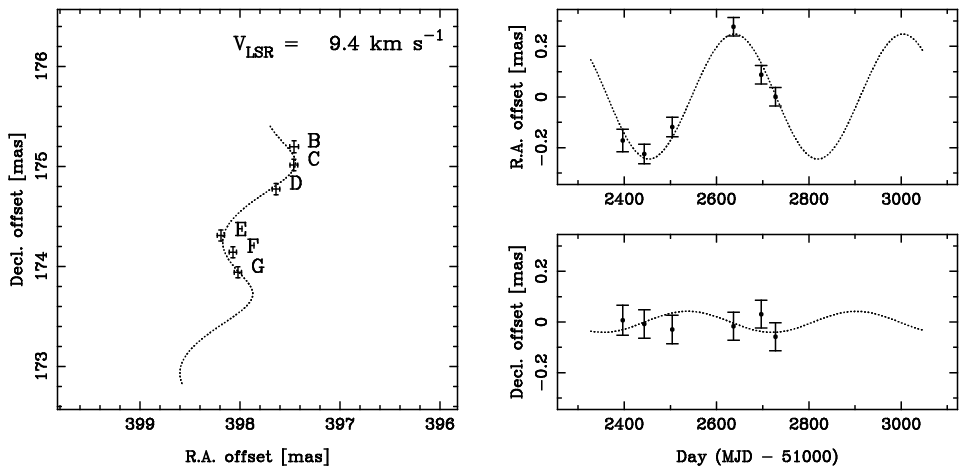} \\
\includegraphics[width=75mm]{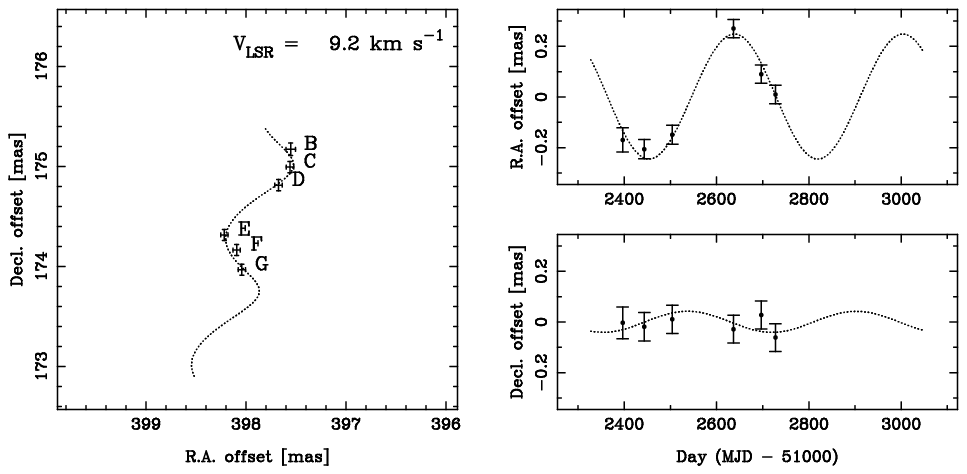}
\includegraphics[width=75mm]{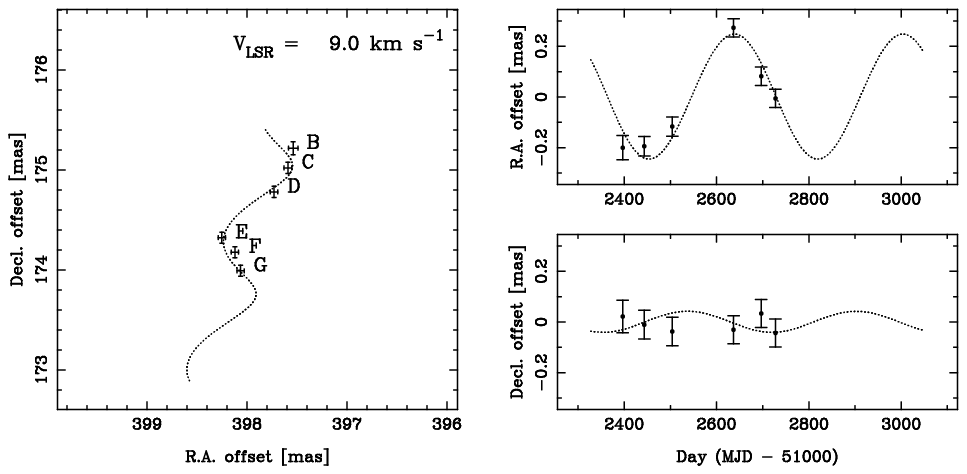} \\
\includegraphics[width=75mm]{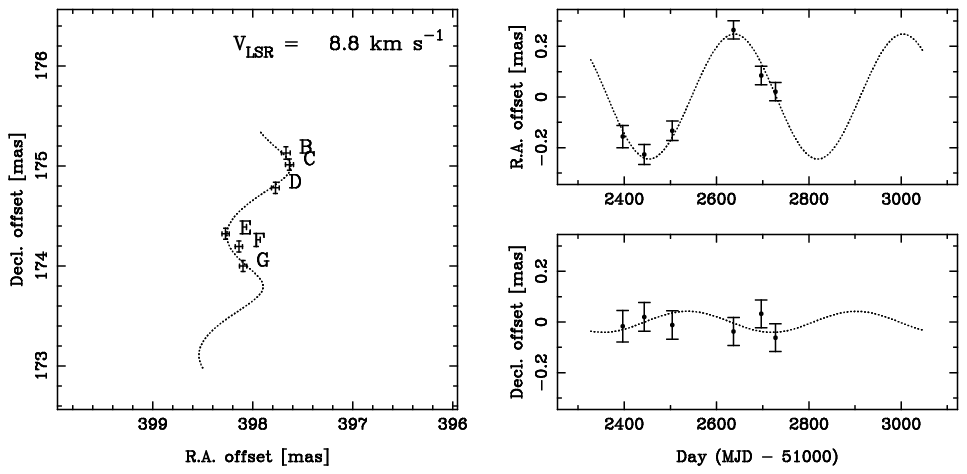}
\includegraphics[width=75mm]{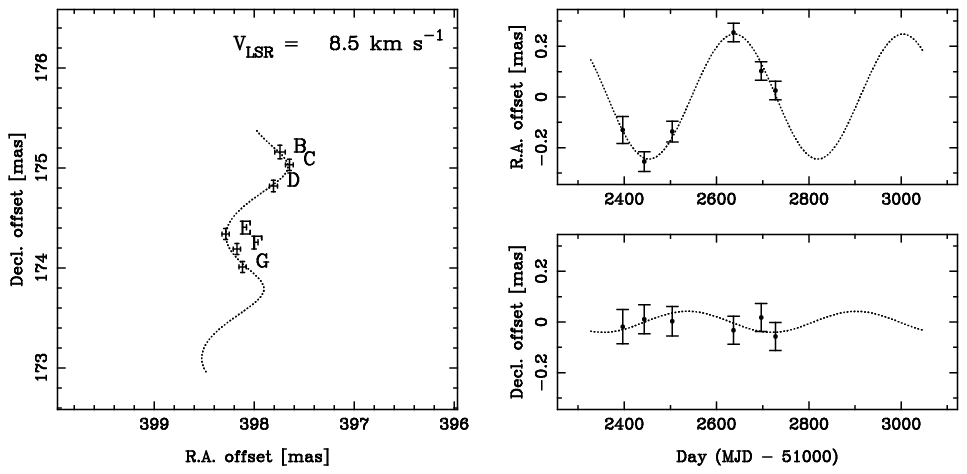} \\
\includegraphics[width=75mm]{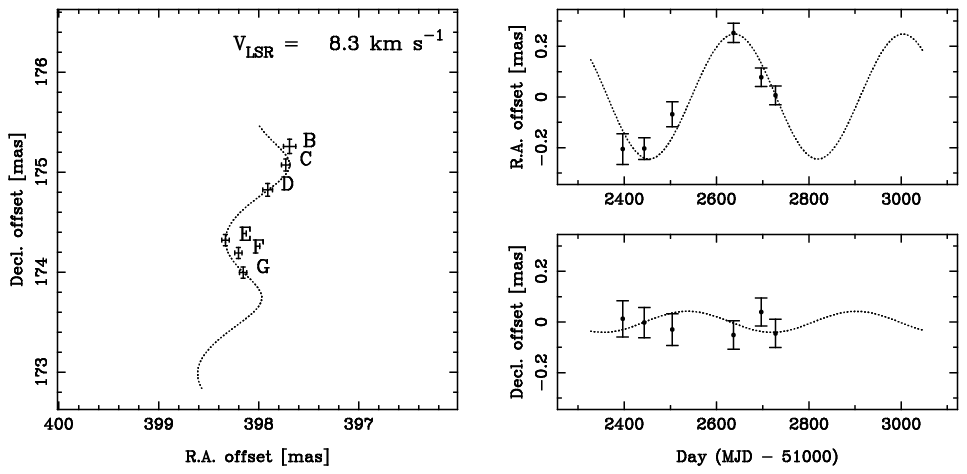}
\includegraphics[width=75mm]{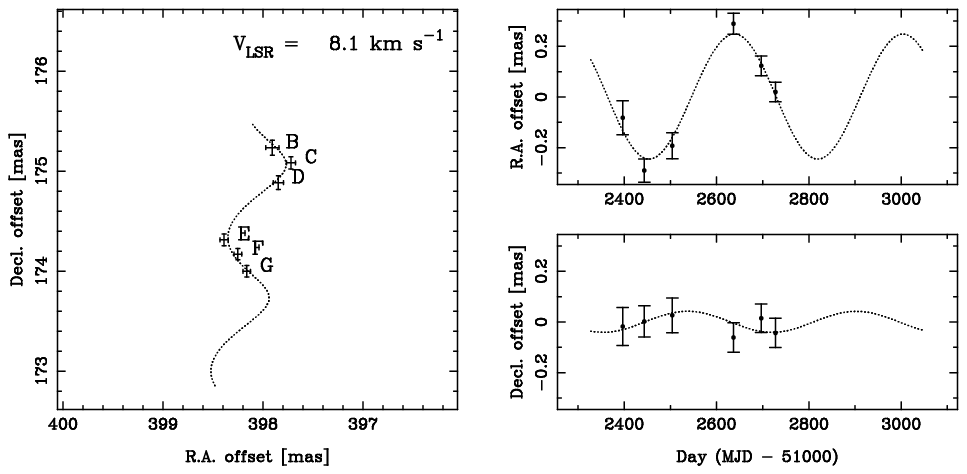}
\end{tabular}
\end{center}
\caption 
{\label{fig:04}
Same as 
Figure~\ref{fig:03}, 
but for the 10 selected maser spots for annual parallax 
analysis for the elliptical Gaussian components obtained 
with AIPS JMFIT.  The positions of the maser spots are 
represented by the crossing points of the error bars in 
right ascension and declination. The common annual 
parallax of 0.247~mas was applied to draw the dotted lines. 
}
\end{figure}
%
%

%
%
\begin{table}
  \begin{center}
  \caption{Epochs of the S269~IRS~2w Astrometric Monitoring Observations}
  \label{tbl:01}
  \begin{tabular}{clcc}
\tableline\tableline
     Epoch & 
     \multicolumn{1}{c}{Date} &  Time Range  \\
                  &          &  (UTC)        \\
\tableline
A\tablenotemark{a}   & 2004 Nov 18  & 12:46--21:25   \\ 
B\tablenotemark{a}   & 2005 Jan 26  & 08:48--17:32    \\ 
C\tablenotemark{a}   & 2005 Mar 14 & 05:18--14:02    \\
D\tablenotemark{a}   & 2005 May 14 &  01:16--10:02   \\
E\tablenotemark{a}    & 2005 Sep 23 & 16:46--01:32    \\
F\tablenotemark{a}    & 2005 Nov 22  & 12:36--21:22    \\
G                                   & 2005 Dec 23  & 10:36--19:22   \\
H\tablenotemark{b}   & 2006 Mar 6    & 05:56--14:42   \\
I\tablenotemark{b}    & 2006 May 13   & 01:26--10:12  \\
J\tablenotemark{b}    & 2006 Jul 22    & 20:57--05:42  \\
K\tablenotemark{b}    & 2006 Aug 7   & 19:56--04:42   \\
L\tablenotemark{b}    & 2006  Sep 7  & 07:56--02:42  \\
M\tablenotemark{b}    & 2006 Oct 16 &   15:26--00:12  \\
N\tablenotemark{b}    & 2006 Nov 10 & 13:26--22:12  \\
\tableline
\tableline
  \end{tabular}
  \tablenotetext{a}{Previously reported by 
H2007 
and 
M2012. 
}
  \tablenotetext{b}{The data were used not for the annual parallax  
  analysis but for the proper motion analysis.}
\end{center}
\end{table}
\clearpage
%
%
\begin{table}
  \begin{center}
  \caption{Phase Tracking Center Positions of the Observed Sources}
  \label{tbl:02}
  \begin{tabular}{lcc}
\tableline\tableline
    
  & Right Ascension (J2000)
  & Declination (J2000) \\
  \tableline
    S269~IRS~2w\tablenotemark{a}
  & $06^{\mathrm{h}} 14^{\mathrm{m}} 37 \fs 03845$
  & $+13^{\circ} 49' 36 \farcs 2200$ \\
    J0613+1306
  & $06^{\mathrm{h}} 13^{\mathrm{m}} 57 \fs 69276$
  & $+13^{\circ} 06' 45 \farcs 4012$ \\
\tableline\tableline
  \end{tabular}
\tablenotetext{a}{
        This position is the origin of S269~IRS~2w images of 
        Figures~\ref{fig:01}--\ref{fig:04}. }
\end{center}
\end{table}
\clearpage
%
%
%
\begin{deluxetable}{cccccccc}
\tabletypesize{\scriptsize}
\tablewidth{0pt}
\tablecaption{
  Positions and Absolute Proper Motions of Water Maser Spots in S269~IRS~2w
  \label{tbl:03}  
}
\tablehead{
\colhead{Spot ID           } & 
\colhead{$\Delta\alpha_{A}\cos{\delta}$\tablenotemark{a}} & 
\colhead{$\Delta\delta_{A}$\tablenotemark{a}} &
\colhead{$V_{\mathrm{LSR}}$} &
\colhead{$F_{\mathrm{max}}$\tablenotemark{b}} &
\colhead{Epoch of} &
\colhead{$\mu_{\alpha}
          \cos{\delta}$    } &
\colhead{$\mu_{\delta}$    } \\
\colhead{(Group ID-       } &
\colhead{(mas)             } &
\colhead{(mas)             } &
\colhead{(km~s$^{-1}$)     } &
\colhead{(Jy~beam$^{-1}$)  } &
\colhead{the peak          } &
\colhead{(mas~yr$^{-1}$)   } &
\colhead{(mas~yr$^{-1}$)   } \\
\colhead{Feature ID)       } &
\colhead{                  } &
\colhead{                  } &
\colhead{                  } &
\colhead{                  } &
\colhead{                  } &
\colhead{                  } &
\colhead{                  } \\
}
\startdata
%
  1(1-a) & $  805.11 \pm     0.03$ & $  470.31 \pm     0.02$ & $    13.6$ & $     2.0$ & I & $   0.203 \pm    0.020$ & $   0.055 \pm    0.016$ \\
  2(1-a) & $  805.30 \pm     0.03$ & $  470.59 \pm     0.02$ & $    13.4$ & $     2.5$ & H & $   0.042 \pm    0.019$ & $  -0.140 \pm    0.015$ \\
  3(2-a) & $  598.22 \pm     0.02$ & $  474.58 \pm     0.01$ & $    20.3$ & $     2.6$ & C & $  -0.875 \pm    0.015$ & $  -0.235 \pm    0.010$ \\
  4(2-a) & $  598.15 \pm     0.01$ & $  474.58 \pm     0.01$ & $    20.1$ & $     8.4$ & D & $  -0.776 \pm    0.010$ & $  -0.225 \pm    0.008$ \\
  5(2-a) & $  598.08 \pm     0.01$ & $  474.57 \pm     0.01$ & $    19.9$ & $    53.9$ & D & $  -0.739 \pm    0.008$ & $  -0.252 \pm    0.007$ \\
  6(2-a) & $  597.92 \pm     0.01$ & $  474.54 \pm     0.01$ & $    19.7$ & $   109.0$ & C & $  -0.738 \pm    0.008$ & $  -0.249 \pm    0.007$ \\
  7(2-a) & $  597.78 \pm     0.01$ & $  474.52 \pm     0.01$ & $    19.5$ & $    82.5$ & C & $  -0.718 \pm    0.008$ & $  -0.246 \pm    0.007$ \\
  8(2-a) & $  597.80 \pm     0.01$ & $  474.51 \pm     0.01$ & $    19.3$ & $    27.8$ & B & $  -0.725 \pm    0.009$ & $  -0.260 \pm    0.007$ \\
  9(2-a) & $  597.80 \pm     0.01$ & $  474.57 \pm     0.01$ & $    19.1$ & $     4.7$ & B & $  -0.747 \pm    0.013$ & $  -0.313 \pm    0.010$ \\
 10(2-a) & $  598.20 \pm     0.02$ & $  474.60 \pm     0.01$ & $    18.9$ & $     1.7$ & C & $  -0.934 \pm    0.028$ & $  -0.312 \pm    0.016$ \\
 11(3-a) & $  383.94 \pm     0.02$ & $  203.53 \pm     0.02$ & $    14.7$ & $     1.1$ & F & $  -0.443 \pm    0.024$ & $   0.976 \pm    0.024$ \\
 12(3-b) & $  384.69 \pm     0.03$ & $  203.86 \pm     0.03$ & $    15.1$ & $     1.3$ & D & $  -1.083 \pm    0.030$ & $   0.700 \pm    0.031$ \\
 13(3-b) & $  384.45 \pm     0.02$ & $  203.88 \pm     0.02$ & $    14.4$ & $     1.7$ & F & $  -0.932 \pm    0.022$ & $   0.622 \pm    0.021$ \\
 14(3-b) & $  384.60 \pm     0.02$ & $  203.78 \pm     0.02$ & $    14.2$ & $     2.7$ & B & $  -1.067 \pm    0.021$ & $   0.732 \pm    0.021$ \\
 15(3-b) & $  384.61 \pm     0.02$ & $  203.81 \pm     0.02$ & $    14.0$ & $     3.7$ & B & $  -1.071 \pm    0.021$ & $   0.700 \pm    0.021$ \\
 16(3-b) & $  384.62 \pm     0.02$ & $  203.78 \pm     0.02$ & $    13.8$ & $     3.3$ & B & $  -1.061 \pm    0.023$ & $   0.705 \pm    0.025$ \\
 17(3-b) & $  384.62 \pm     0.02$ & $  203.82 \pm     0.02$ & $    13.6$ & $     2.2$ & B & $  -1.034 \pm    0.025$ & $   0.626 \pm    0.026$ \\
 18(3-b) & $  384.66 \pm     0.02$ & $  203.80 \pm     0.02$ & $    13.4$ & $     1.8$ & G & $  -1.087 \pm    0.025$ & $   0.596 \pm    0.026$ \\
 19(3-c) & $  404.40 \pm     0.02$ & $  173.75 \pm     0.02$ & $    12.3$ & $     4.4$ & M & $   0.348 \pm    0.011$ & $  -1.260 \pm    0.010$ \\
 20(3-c) & $  404.21 \pm     0.02$ & $  173.50 \pm     0.02$ & $    12.1$ & $     3.0$ & M & $   0.453 \pm    0.012$ & $  -1.121 \pm    0.011$ \\
 21(3-d) & $  403.50 \pm     0.02$ & $  173.92 \pm     0.02$ & $    13.6$ & $     5.2$ & N & $   0.280 \pm    0.011$ & $  -1.223 \pm    0.010$ \\
 22(3-e) & $  403.59 \pm     0.02$ & $  174.26 \pm     0.02$ & $    12.5$ & $     3.2$ & M & $   0.785 \pm    0.012$ & $  -1.535 \pm    0.011$ \\
 23(3-f) & $  403.14 \pm     0.02$ & $  174.73 \pm     0.02$ & $    13.8$ & $     4.4$ & G & $   0.083 \pm    0.018$ & $  -1.732 \pm    0.018$ \\
 24(3-f) & $  403.29 \pm     0.02$ & $  174.47 \pm     0.02$ & $    13.6$ & $     2.4$ & G & $  -0.035 \pm    0.020$ & $  -1.483 \pm    0.019$ \\
 25(3-f) & $  403.18 \pm     0.02$ & $  174.50 \pm     0.02$ & $    13.4$ & $     2.5$ & D & $   0.186 \pm    0.024$ & $  -1.512 \pm    0.022$ \\
 26(3-g) & $  402.61 \pm     0.02$ & $  174.49 \pm     0.02$ & $    14.7$ & $     4.7$ & H & $   0.902 \pm    0.011$ & $  -1.627 \pm    0.009$ \\
 27(3-g) & $  402.18 \pm     0.01$ & $  174.51 \pm     0.01$ & $    14.4$ & $     7.0$ & L & $   1.113 \pm    0.009$ & $  -1.613 \pm    0.008$ \\
 28(3-g) & $  402.23 \pm     0.01$ & $  174.49 \pm     0.01$ & $    14.2$ & $     9.7$ & L & $   1.057 \pm    0.008$ & $  -1.584 \pm    0.007$ \\
 29(3-g) & $  402.13 \pm     0.01$ & $  174.63 \pm     0.01$ & $    14.0$ & $     9.2$ & M & $   1.088 \pm    0.008$ & $  -1.655 \pm    0.007$ \\
 30(3-h) & $  401.99 \pm     0.02$ & $  174.78 \pm     0.02$ & $    12.3$ & $     8.8$ & D & $   1.887 \pm    0.045$ & $  -1.438 \pm    0.043$ \\
 31(3-h) & $  402.27 \pm     0.02$ & $  174.68 \pm     0.02$ & $    12.1$ & $     7.8$ & C & $   1.559 \pm    0.074$ & $  -1.016 \pm    0.072$ \\
 32(3-h) & $  402.26 \pm     0.02$ & $  174.85 \pm     0.02$ & $    11.9$ & $     6.4$ & C & $   1.402 \pm    0.050$ & $  -1.520 \pm    0.048$ \\
 33(3-h) & $  402.36 \pm     0.02$ & $  174.69 \pm     0.02$ & $    11.7$ & $     3.8$ & C & $   1.226 \pm    0.078$ & $  -0.950 \pm    0.076$ \\
 34(3-h) & $  402.20 \pm     0.03$ & $  174.71 \pm     0.03$ & $    11.5$ & $     2.0$ & C & $   1.812 \pm    0.093$ & $  -0.971 \pm    0.090$ \\
 35(3-i) & $  401.29 \pm     0.02$ & $  174.74 \pm     0.02$ & $    13.8$ & $     7.7$ & M & $   1.497 \pm    0.010$ & $  -1.692 \pm    0.009$ \\
 36(3-j) & $  400.80 \pm     0.02$ & $  174.81 \pm     0.02$ & $    12.8$ & $     1.7$ & A & $   2.886 \pm    0.109$ & $  -1.285 \pm    0.094$ \\
 37(3-j) & $  400.75 \pm     0.02$ & $  174.78 \pm     0.02$ & $    12.5$ & $     2.2$ & A & $   2.799 \pm    0.099$ & $  -0.946 \pm    0.088$ \\
 38(3-k) & $  397.29 \pm     0.02$ & $  175.05 \pm     0.01$ & $    10.9$ & $     3.3$ & F & $   0.208 \pm    0.014$ & $  -0.950 \pm    0.014$ \\
 39(3-k) & $  397.25 \pm     0.01$ & $  175.27 \pm     0.01$ & $    10.7$ & $     7.7$ & F & $   0.242 \pm    0.011$ & $  -1.148 \pm    0.010$ \\
 40(3-k) & $  397.34 \pm     0.01$ & $  175.37 \pm     0.01$ & $    10.4$ & $    10.9$ & F & $   0.180 \pm    0.011$ & $  -1.203 \pm    0.009$ \\
 41(3-k) & $  397.45 \pm     0.01$ & $  175.28 \pm     0.01$ & $    10.2$ & $     9.7$ & H & $   0.110 \pm    0.010$ & $  -1.107 \pm    0.009$ \\
 42(3-k) & $  397.53 \pm     0.01$ & $  175.39 \pm     0.01$ & $    10.0$ & $     6.0$ & H & $   0.110 \pm    0.011$ & $  -1.193 \pm    0.010$ \\
 43(3-k) & $  397.65 \pm     0.01$ & $  175.32 \pm     0.01$ & $     9.8$ & $     4.6$ & F & $   0.129 \pm    0.013$ & $  -1.138 \pm    0.012$ \\
 44(3-k) & $  397.71 \pm     0.01$ & $  175.35 \pm     0.01$ & $     9.6$ & $     7.8$ & F & $   0.159 \pm    0.013$ & $  -1.196 \pm    0.012$ \\
 45(3-k) & $  397.75 \pm     0.01$ & $  175.40 \pm     0.01$ & $     9.4$ & $    13.3$ & F & $   0.167 \pm    0.013$ & $  -1.240 \pm    0.012$ \\
 46(3-k) & $  397.84 \pm     0.01$ & $  175.40 \pm     0.01$ & $     9.2$ & $    18.0$ & F & $   0.125 \pm    0.013$ & $  -1.243 \pm    0.012$ \\
 47(3-k) & $  397.86 \pm     0.01$ & $  175.41 \pm     0.01$ & $     9.0$ & $    18.9$ & F & $   0.133 \pm    0.013$ & $  -1.232 \pm    0.012$ \\
 48(3-k) & $  397.93 \pm     0.01$ & $  175.38 \pm     0.01$ & $     8.8$ & $    16.6$ & F & $   0.084 \pm    0.013$ & $  -1.203 \pm    0.012$ \\
 49(3-k) & $  398.00 \pm     0.01$ & $  175.40 \pm     0.01$ & $     8.5$ & $    10.8$ & F & $   0.018 \pm    0.013$ & $  -1.198 \pm    0.013$ \\
 50(3-k) & $  398.02 \pm     0.01$ & $  175.48 \pm     0.01$ & $     8.3$ & $     5.4$ & F & $   0.036 \pm    0.014$ & $  -1.284 \pm    0.013$ \\
 51(3-k) & $  398.07 \pm     0.01$ & $  175.49 \pm     0.01$ & $     8.1$ & $     2.7$ & F & $   0.022 \pm    0.016$ & $  -1.300 \pm    0.015$ \\
 52(3-k) & $  397.93 \pm     0.02$ & $  175.52 \pm     0.02$ & $     7.9$ & $     1.6$ & F & $   0.175 \pm    0.020$ & $  -1.289 \pm    0.019$ \\
 53(3-k) & $  397.49 \pm     0.03$ & $  175.78 \pm     0.03$ & $     7.7$ & $     1.0$ & F & $   0.637 \pm    0.025$ & $  -1.554 \pm    0.026$ \\
 54(3-l) & $  398.03 \pm     0.03$ & $  175.70 \pm     0.04$ & $     4.5$ & $     1.6$ & B & $   0.274 \pm    0.134$ & $  -0.539 \pm    0.172$ \\
 55(3-l) & $  398.19 \pm     0.03$ & $  175.79 \pm     0.04$ & $     4.3$ & $     1.3$ & C & $  -0.429 \pm    0.125$ & $  -0.753 \pm    0.167$ \\
 56(3-l) & $  398.22 \pm     0.04$ & $  176.05 \pm     0.05$ & $     4.1$ & $     1.0$ & C & $  -0.358 \pm    0.146$ & $  -1.633 \pm    0.177$ \\
 57(4-a) & $  201.42 \pm     0.02$ & $  -55.40 \pm     0.02$ & $    20.1$ & $     3.6$ & F & $   1.076 \pm    0.020$ & $  -2.254 \pm    0.021$ \\
 58(4-a) & $  202.09 \pm     0.02$ & $  -56.34 \pm     0.02$ & $    19.3$ & $     7.6$ & F & $   0.423 \pm    0.015$ & $  -1.319 \pm    0.015$ \\
 59(4-a) & $  202.09 \pm     0.02$ & $  -56.43 \pm     0.02$ & $    19.1$ & $     2.9$ & E & $   0.433 \pm    0.018$ & $  -1.245 \pm    0.018$ \\
 60(4-b) & $  204.94 \pm     0.02$ & $  -55.10 \pm     0.02$ & $    18.9$ & $     3.4$ & C & $   1.433 \pm    0.080$ & $  -0.904 \pm    0.080$ \\
 61(4-c) & $  182.69 \pm     0.02$ & $  -65.34 \pm     0.02$ & $    18.4$ & $     5.3$ & C & $   1.193 \pm    0.072$ & $  -1.217 \pm    0.072$ \\
 62(4-c) & $  182.61 \pm     0.02$ & $  -65.29 \pm     0.02$ & $    18.2$ & $    15.4$ & C & $   1.473 \pm    0.069$ & $  -1.371 \pm    0.069$ \\
 63(4-c) & $  182.66 \pm     0.02$ & $  -65.36 \pm     0.02$ & $    18.0$ & $    18.3$ & C & $   1.322 \pm    0.071$ & $  -1.159 \pm    0.071$ \\
 64(4-d) & $  178.12 \pm     0.02$ & $  -67.43 \pm     0.02$ & $    17.4$ & $     6.9$ & B & $   0.228 \pm    0.100$ & $  -1.602 \pm    0.094$ \\
 65(4-d) & $  178.12 \pm     0.02$ & $  -67.45 \pm     0.02$ & $    17.2$ & $    16.7$ & B & $   0.236 \pm    0.082$ & $  -1.415 \pm    0.083$ \\
 66(4-d) & $  178.07 \pm     0.02$ & $  -67.43 \pm     0.02$ & $    17.0$ & $    31.3$ & B & $   0.497 \pm    0.083$ & $  -1.518 \pm    0.080$ \\
 67(4-d) & $  178.05 \pm     0.02$ & $  -67.45 \pm     0.02$ & $    16.8$ & $    40.1$ & B & $   0.656 \pm    0.086$ & $  -1.382 \pm    0.083$ \\
 68(4-d) & $  177.94 \pm     0.02$ & $  -67.22 \pm     0.02$ & $    16.6$ & $    31.5$ & B & $   1.229 \pm    0.097$ & $  -2.568 \pm    0.100$ \\
 69(4-e) & $  167.24 \pm     0.02$ & $  -73.02 \pm     0.02$ & $    20.6$ & $     3.3$ & C & $   1.446 \pm    0.080$ & $  -1.443 \pm    0.080$ \\
 70(4-e) & $  167.37 \pm     0.02$ & $  -72.98 \pm     0.02$ & $    20.3$ & $     7.4$ & C & $   0.974 \pm    0.072$ & $  -1.567 \pm    0.071$ \\
 71(4-e) & $  167.23 \pm     0.02$ & $  -72.92 \pm     0.02$ & $    20.1$ & $    12.0$ & C & $   1.380 \pm    0.070$ & $  -1.781 \pm    0.070$ \\
 72(4-f) & $  157.51 \pm     0.02$ & $  -75.97 \pm     0.02$ & $    20.6$ & $     1.4$ & F & $  -1.879 \pm    0.023$ & $  -3.355 \pm    0.023$ \\
 73(4-f) & $  156.31 \pm     0.02$ & $  -76.55 \pm     0.02$ & $    20.3$ & $     2.9$ & G & $  -0.866 \pm    0.021$ & $  -2.818 \pm    0.020$ \\
 74(5-a) & $  -98.19 \pm     0.02$ & $ -199.49 \pm     0.02$ & $    21.0$ & $     3.2$ & M & $   0.439 \pm    0.011$ & $  -2.741 \pm    0.010$ \\
 75(5-b) & $  -97.61 \pm     0.02$ & $ -200.04 \pm     0.02$ & $    20.8$ & $     4.1$ & K & $   0.126 \pm    0.011$ & $  -2.413 \pm    0.010$ \\
 76(5-b) & $  -97.35 \pm     0.02$ & $ -199.88 \pm     0.02$ & $    20.6$ & $     9.1$ & K & $   0.001 \pm    0.010$ & $  -2.515 \pm    0.009$ \\
 77(5-c) & $  -96.56 \pm     0.02$ & $ -200.49 \pm     0.02$ & $    19.1$ & $   596.0$ & M & $  -0.441 \pm    0.010$ & $  -2.174 \pm    0.009$ \\
 78(5-d) & $  -97.14 \pm     0.02$ & $ -201.38 \pm     0.02$ & $    18.9$ & $   316.4$ & M & $  -0.136 \pm    0.011$ & $  -1.698 \pm    0.010$ \\
 79(5-e) & $  -98.78 \pm     0.02$ & $ -200.98 \pm     0.01$ & $    20.3$ & $    21.2$ & K & $   0.799 \pm    0.009$ & $  -1.927 \pm    0.008$ \\
 80(5-e) & $  -98.86 \pm     0.02$ & $ -201.02 \pm     0.01$ & $    20.1$ & $    44.9$ & K & $   0.844 \pm    0.009$ & $  -1.904 \pm    0.008$ \\
 81(5-e) & $  -98.78 \pm     0.02$ & $ -201.00 \pm     0.01$ & $    19.9$ & $    67.7$ & M & $   0.793 \pm    0.009$ & $  -1.916 \pm    0.008$ \\
 82(5-e) & $  -98.73 \pm     0.02$ & $ -201.08 \pm     0.01$ & $    19.7$ & $   236.3$ & M & $   0.756 \pm    0.009$ & $  -1.874 \pm    0.008$ \\
 83(5-e) & $  -98.72 \pm     0.02$ & $ -201.11 \pm     0.02$ & $    19.5$ & $   485.6$ & M & $   0.743 \pm    0.010$ & $  -1.851 \pm    0.008$ \\
 84(5-e) & $  -98.72 \pm     0.02$ & $ -201.26 \pm     0.02$ & $    19.3$ & $   652.4$ & M & $   0.728 \pm    0.010$ & $  -1.768 \pm    0.009$ \\
 85(6-a) & $ -476.60 \pm     0.02$ & $ -198.00 \pm     0.02$ & $    19.1$ & $     1.8$ & E & $  -0.316 \pm    0.017$ & $  -0.853 \pm    0.017$ \\
 86(6-a) & $ -476.38 \pm     0.02$ & $ -197.39 \pm     0.02$ & $    18.9$ & $     1.0$ & E & $  -0.577 \pm    0.023$ & $  -1.398 \pm    0.022$ \\
 87(6-b) & $ -550.10 \pm     0.02$ & $ -262.55 \pm     0.02$ & $    18.4$ & $     2.4$ & E & $  -0.598 \pm    0.018$ & $  -1.041 \pm    0.018$ \\
 88(6-b) & $ -550.39 \pm     0.02$ & $ -262.44 \pm     0.02$ & $    18.2$ & $     2.9$ & E & $  -0.328 \pm    0.018$ & $  -1.144 \pm    0.019$ \\
 89(6-b) & $ -550.38 \pm     0.02$ & $ -262.62 \pm     0.02$ & $    18.0$ & $     1.4$ & E & $  -0.316 \pm    0.022$ & $  -0.970 \pm    0.021$ \\
 90(6-c) & $ -666.42 \pm     0.03$ & $ -297.23 \pm     0.03$ & $    18.7$ & $     1.7$ & B & $   0.968 \pm    0.119$ & $  -0.311 \pm    0.108$ \\
\enddata
\tablenotetext{a}{
  The positions calculated at epoch~A without the annual parallax modulation. 
  The origin is located at the phase tracking center position of S269~IRS~2w as 
  listed in Table~\ref{tbl:02}. 
}
\tablenotetext{b}{
  The maximum value of the peak flux density during the monitor program. 
}
\end{deluxetable}
\clearpage
%
%



\clearpage




\end{document}